\documentclass[12pt]{article}
\usepackage{epsfig}

\def\beq{\begin{equation}}
\def\eeq#1{\label{#1}\end{equation}}
\def\eeqn{\end{equation}}

\def\beqa{\begin{eqnarray}}
\def\eeqa#1{\label{#1}\end{eqnarray}}
\def\eeqan{\end{eqnarray}}

\let\bar=\overbar

\def\Dslash{\not{\hbox{\kern-4pt $D$}}}
\def\dslash{\not{\hbox{\kern-2pt $\del$}}}

\def\msb{{\bar{\ssstyle M \kern -1pt S}}}

\def\Title#1{\begin{center} {\Large {\bf #1} } \end{center}}
\def\beq{\begin{equation}}
\def\eeq{\end{equation}}
\def\beqa{\begin{eqnarray}}
\def\eeqa{\end{eqnarray}}
\def\ss{\scriptscriptstyle}

\newcommand{\AD}[1]{$\overline{\mbox{D~\,}}\!\!\!$#1}
\def\ss{\scriptscriptstyle}

\begin{document}

\Title{Electromagnetic scattering of vector mesons in the Sakai-Sugimoto model}

\bigskip\bigskip


\begin{raggedright}  

{\it C. A. Ballon Bayona 
\\
 Centre for Particle Theory, University of Durham, \\
Science Laboratories, South Road, Durham DH1 3LE -- United Kingdom \\
email: c.a.m.ballonbayona@durham.ac.uk}

\bigskip

{\it Henrique Boschi-Filho, Nelson R. F. Braga, Marcus A. C. Torres \\
Instituto de F\'{\i}sica, Universidade Federal do Rio de Janeiro, \\
Caixa Postal 68528, RJ 21941-972 -- Brazil\\
emails: boschi@if.ufrj.br, braga@if.ufrj.br, mtorres@if.ufrj.br} 

\bigskip\bigskip
\end{raggedright}

\abstract{In this proceeding we review some of our recent results for the vector meson   electromagnetic form factors 
and structure functions in the Sakai-Sugimoto model. 
The latter is a string model that describes 
 many features of the non-perturbative regime of Quantum Chromodynamics in the large $N_c$ limit.}

\section{Introduction}

It was shown by 't Hooft in 1973 \cite{'tHooft:1973jz}
 that considering the large $N_c$ limit of Quantum Chromodynamics (QCD) the amplitudes simplify and we can treat $1/N_c$ as a parameter for perturbative expansions. Moreover, the Feynman diagrams can be redrawn as surfaces with different topologies so that the $1/N_c$ expansion 
is interpreted as an expansion in the genus of the surfaces. 
The 't Hooft approach is known as large $N_c$ QCD and the genus expansion is similar to the one that arises in String Theory suggesting  a  duality between quantum field theories and string theories. 

\medskip 

The first concrete duality between quantum field theories and string theories was proposed by  Maldacena  in 1997 \cite{Maldacena:1997re}. 
A careful analysis of brane embeddings in SuperString/M Theory led him to conjecture a correspondence between quantum field theories with conformal symmetry (CFT) in $D$ 
spacetime dimensions and Superstring/M theory compactifications on a $D + 1$ Anti-de-Sitter (AdS) spacetime.  The Maldacena duality, known as the AdS/CFT correspondence, 
relates the strong (weak)  coupling regime of a quantum field theory to the weak 
(strong) coupling regime of a Superstring/M theory. 

\medskip 

Inspired by the AdS/CFT correspondence, several quantum field theories that have string duals have been investigated. 
Of special interest is the study of string models dual to strongly coupled theories similar to QCD. 
This intense area of research is known as AdS/QCD (for a review see \cite{Peeters:2007ab,Erdmenger:2007cm,Gubser:2009md}). 
This is a powerful approach to non-perturbative QCD because many results do not depend on the particular string model. 
There is one string model, though, that has a field content very similar to large $N_c$ QCD in the regime of large distances and massless quarks. This is the Sakai-Sugimoto model (2004) 
\cite{Sakai:2004cn} that realizes confinement and chiral symmetry breaking.

\medskip

The Sakai-Sugimoto model predicts at large distances an effective four-dimensional lagrangian that includes the non-linear sigma model with a Skyrme interaction term. The latter is a very 
successful effective model for large $N_c$ QCD at large distances. On top of that, the Sakai-Sugimoto model allows the description of vector mesons which are a key ingredient in hadronic 
interactions. This is because of the experimentally observed vector meson dominance \cite{Sakurai:1969} which is the decomposition of photon-hadron interaction 
into vector meson exchange. All these features turn the Sakai-Sugimoto model into a good candidate for predicting scattering cross sections in non-perturbative QCD.

\medskip 

We review in this proceeding the calculation of vector meson electromagnetic form factors \cite{BallonBayona:2009ar} and structure functions \cite{BallonBayona:2010ae}
in the Sakai-Sugimoto model. We begin this review recalling in section 2 the basic langauge of electromagnetic scattering of hadrons. Then we review in section 3 
the Sakai-Sugimoto model and the property of vector meson dominance. In section 4 we describe the calculation of vector meson electromagnetic form factors and structure
functions. For simplicity,  we focus on the case where  the initial state is a $\rho$ meson. We show how to extract from the Sakai-Sugimoto model a generalized 
electromagnetic form factor that includes elastic and non-elastic scattering. 
As for the structure functions, we consider only processes where one vector meson resonance is produced in the final state. 
We finish the review with some conclusions and perspectives of future developments.

\section{Electromagnetic scattering of hadrons}

Consider the electromagnetic scattering of a lepton $\ell^-$ with momentum $k$ and a hadron $H$ with momentum $p$ :
\beqa
\ell^-(k) \,+\, H(p) \,\to \, \ell^-(k') + X \,. \label{eHeX}
\eeqa
In this process the lepton and hadron exchange a virtual photon $\gamma^*$ with momentum $q^\mu \,=\, k^\mu - k'^\mu$.
The $X$ symbol represents a hadronic final state that is  (or is not )  measured when the scattering is 
exclusive (inclusive). We will focus on the case where $X$ represent one hadronic resonance although in general $X$ may represent 
two or more particles.  The kinematical variables relevant for this process are 

\begin{itemize}
 \item The mass of the initial hadron : \quad $M = \sqrt{-p^2}$ \, ,
 \item  The virtuality : \quad $q^2 \,=\, - q_0^2 + \vec{q}^2 > 0 $ \, , 
 \item The Bjorken variable : \quad $ x \,=\, - q^2 / (2 p \cdot q) $ \, ,
 \item The photon-hadron c.m. energy : \quad $W  \,=\, \sqrt{- (p + q)^2}$ \,. 
\end{itemize} 
The physical region of scattering is $  0 < x  \le 1$ and elastic scattering corresponds to $ x = 1$. 
Note that we are working in the signature $(-,+,+,+)$.

\medskip

Now consider Deep Inelastic Scattering (DIS) which is the inclusive electromagnetic scattering of a lepton and a hadron. Using the Feynman rules for the process (\ref{eHeX}), the DIS differential
 cross section (for the unpolarized case) can be written as 
\beqa
d^2 \sigma \,=\, \left ( \frac{1}{4 M E} \right ) \frac{d^3 \vec{k}}{(2 \pi)^3 2 E'} \, \frac{e^4}{q^4} (4 \pi)\, L_{\mu \nu} \, W^{\mu \nu} \, ,
\eeqa 
where 
\beqa
L_{\mu \nu} &\,=\,& \frac12 \sum_{\sigma,\sigma'} \bar u (k,\sigma) \gamma_\mu u(k', \sigma') \, \bar u (k',\sigma') \gamma_\nu u(k, \sigma) \, , \cr
W^{\mu \nu} &\,=\,& \frac{1}{4\pi (2 s_H + 1)}  \sum_{\sigma_H} \sum_X (2 \pi)^4 \delta^4( p + q - p_X) \langle p, \sigma_H  \vert J^\mu (0) \vert X \rangle  \cr 
&\times& \langle X \vert J^\nu(0) \vert p, \sigma_H \rangle \, .
\eeqa 
The tensor $L_{\mu \nu}$  carries the information of the initial and final leptons (leptonic tensor) while the
tensor $W^{\mu \nu}$ describes the response of a hadron H to an electromagnetic current $J^\nu(x)$ (hadronic tensor).  
Using current conservation and relevant QCD symmetries like parity, time reversal and Lorentz invariance we can decompose the hadronic tensor as
\beqa \label{eq:strfun}
W^{\mu\nu} = F_1 (x,q^2)  \Big( \eta^{\mu\nu} - \frac{q^\mu q^\nu}{q^2}  \Big) 
\,+\,\frac{2x}{q^2} F_2 (x,q^2)  \Big( p^\mu +  \frac{q^\mu}{2x}  \Big) 
\Big( p^\nu \,+ \, \frac{q^\nu}{2x} \, \Big) \, , \label{HadTensor}
\eeqa
where $F_1 (x,q^2)$ and $F_2 (x,q^2)$ are known as the {\it structure functions} of the hadron. 

\medskip 

In the hadronic tensor each final state $X$ contributes through the current matrix element $\langle p, \sigma_H  \vert J^\mu_H(0) \vert X \rangle$. 
In the case where the final state is  one hadron with  the same quantum numbers  as the initial hadron but a different mass the current matrix element can be decomposed as
\beqa
\langle p, n , \sigma_H  \vert J^\mu (0) \vert p + q , n_X , \sigma_X \rangle \,=\, \sum_i \Gamma_i^\mu (p,q) \,  F^i_{n,n_X}(q^2) \, ,
\eeqa
where $n$ ($n_X$) is an index associated to the mass of the initial (final) hadron. The functions $F^i_{n,n_X}(q^2)$  are the {\it generalized electromagnetic form factors}
that include the elastic ($n = n_X$) and transition ($n \ne n_X$) form factors. 

\medskip 

As a example, let's consider the elastic $\rho$ meson form factor where $n_X = n = 1$. In that case the current can be decomposed as 
\beqa
\langle p, \epsilon \vert J^{\mu }(0) \vert   p + q, \epsilon' \rangle 
&=&   \epsilon \cdot \epsilon' (2p+q)^\mu F_1 ( q^2 )  
+ \left[ \epsilon^\mu \epsilon'\cdot q - {\epsilon'}^\mu \epsilon \cdot q  \right] \Big [  F_1 (q^2) \cr
&+&  F_2(q^2) \Big ] - \frac{q\cdot \epsilon' q \cdot \epsilon }{M^2}  (2p+q)^\mu  F_3 (q^2) \, . \label{FormDecomp}
\eeqa

\section{The Sakai-Sugimoto model}

In the  regime of low momentum transfer ($\sqrt{q^2}$ lower than some $GeVs$), soft (non-perturbative)
processes become dominant in hadronic scattering. 
Lattice QCD (the numerical approach of substituting the continuous spacetime by a lattice of points) is not very useful here because of 
time dependence in scattering processes. Effective models are very useful but they don't succeed completely when comparing with
experimental results. As stressed in the introduction, the Sakai-Sugimoto model provides a new insight into the problem 
of hadronic scattering in the non-perturbative regime. This model consists in the introduction of $N_f$ D8-\AD8
branes into a set of $N_c$ D4 branes in the limit of large $N_c$ with fixed $N_f$. 
This limit allows a supergravity description and can be interpreted as the quenching limit in QCD. 
The Sakai-Sugimoto model is the first string model that realizes confinement and chiral symmetry breaking. Below we describe this model 
in some detail. 

\medskip 

Consider a set of $N_c$ coincident D4-branes with a compact spatial direction in type IIA Supergravity \cite{Witten:1998zw}. They generate a background 
composed by a metric, dilaton and four-form :
\beqa
ds^2 &=&  \frac{u^{\ss 3/2}}{R^{\ss 3/2}} \left [ \eta_{\mu \nu} dx^\mu dx^\nu +  f(u) d\tau^2  \right ]
+  \frac{R^{\ss 3/2}}{u^{\ss 3/2}}  \frac{d u^2}{f(u)} + R^{\ss 3/2} u^{\ss 1/2} d \Omega_4^2   \, , \cr
f(u) &=& 1 - \frac{u_{\Lambda}^3}{u^3} \quad , \quad  e^\phi = g_s   \frac{u^{3/4}}{R^{3/4}} \quad , \quad F_4 = \frac{(2 \pi l_s)^3 N_c}{V_{\ss S^4}} \epsilon_4 \,  ,
\eeqa
where $R = (\pi g_s N_c)^{1/3} \sqrt{\alpha'}$. The $\tau$ coordinate is compact with  period 
\beqa
\delta \tau = \frac{4 \pi}{3} \frac{R^{3/2}}{u_{\Lambda}^{1/2}} \equiv \frac{2 \pi}{M_\Lambda} \,,
\eeqa 
where $M_\Lambda$ is a 4-d mass scale. This compactification is introduced as a mechanism of supersymmetry breaking and confinement. 
Imposing anti-periodic conditions for the fermionic states we get at low energies a four-dimensional 
non-supersymmetric strongly coupled $U(N_c)$ theory at large $N_c$ with 't Hooft constant  given by 
\beq
\lambda = g_{YM}^2 N_c = (2 \pi M_\Lambda) \,  g_s N_c \, l_s \,.
\eeq

It is convenient to introduce use a pair of dimensionless coordinates $y$ and $z$ defined by the relations
\begin{equation}
u = u_{\Lambda} \left (1 + y^2 + z^2 \right )^{1/3} \equiv u_{\Lambda}
K_{y,z}^{1/3} \quad , \quad \tau = \frac{\delta \tau}{2 \pi} \arctan
\left (\frac{z}{y} \right) \,.
\end{equation} 
In terms of these coordinates the metric takes the form
\beqa
{\rm d}s^2 &=& u_{\Lambda}^{3/2} R^{-3/2}
K_{y,z}^{1/2} \, \eta_{\mu \nu} {\rm d}x^\mu {\rm d}x^\nu \,+\, \frac49 R^{3/2}
u_{\Lambda}^{1/2} \frac{K_{y,z}^{-5/6}}{y^2 + z^2} \Big [ (z^2 + y^2
K_{y,z}^{1/3}){\rm d}z^2 \cr
&+& (y^2 + z^2 K_{y,z}^{1/3} ) {\rm d}y^2 + 2 y z (1
- K_{y,z}^{1/3}) {\rm d}y {\rm d}z \Big ] + R^{3/2} U_{\ss KK}^{1/2} K_{y,z}^{1/6}
{\rm d}\Omega_4^2 \,.  
\eeqa

\medskip 

Now consider $N_f$ coincident D8-\AD8 probe branes living in the background generated by the $N_c$ D4-branes. The probe approximation is guaranteed by the condition $N_f \ll N_c$.
The $N_f$ D8 branes bring quark degrees of freedom as fundamental strings extending from the D4 branes to the D8 branes. 
The dynamic of the D8 and \AD8 branes is dictated  by the DBI action. It turns out that the solution to the DBI equations merge the D8 and \AD8 branes 
in the infrared region (small $u$). This is a geometrical realization of chiral symmetry breaking  $U(N_f) \times U(N_f) \to U(N_f)$. In the simplest case 
the solution is just $y=0$ (antipodal solution) and the induced D8-\AD8 metric takes the format
\beqa
 {\rm d}s_{D8} &=& u_{\Lambda}^{3/2} R^{-3/2} K_z^{1/2} \, \eta_{\mu \nu} {\rm d}x^\mu
{\rm d}x^\nu \,+\, \frac49 R^{3/2} u_{\Lambda}^{1/2} K_z^{-5/6} {\rm d}z^2  \cr 
&+& R^{3/2} u_{\Lambda}^{1/2} K_z^{1/6} {\rm d}\Omega_4^2 \, , 
\eeqa
where $K_z = 1 + z^2$. Using the DBI action we can describe 9-d gauge field fluctuations in the D8-\AD8 branes. Considering 
small gauge field fluctuations $A_\mu,A_z$ depending only in $x^\mu$ and $z$ directions the DBI action reduces to a 
Yang-Mills action that can be integrated in $S^4$ leading to
\beqa
S_{YM} = -  \kappa \int d^4 x d z {\rm Tr} \left [ \frac12 K_z^{-1/3} \eta^{\mu \rho} \eta^{\nu \sigma}  F_{\mu \nu}  F_{\rho \sigma} 
+  M_\Lambda^2 \, K_z \eta^{\mu \nu}  F_{\mu z}  F_{\nu z}  \right ] \label{5dYM}
\eeqa
where $\kappa = \lambda N_c /(216 \pi^3)$. The gauge field $A_\mu$ can be expanded, in the $A_z=0$ gauge, as
\beqa
A_{\mu} (x, z) =  \hat {\cal V}_\mu (x)  + \hat {\cal A}_\mu (x)  \psi_0 ( z) 
+ \sum_{n=1}^{\infty} \left [ v_\mu^n (x) \psi_{2n-1} (z) +  a_\mu^n (x) \psi_{2n} (z) \right ]  \label{KKexpansion} 
\eeqa
where 
\beqa
\hat {\cal V}_\mu (x)&=& \frac12  U^{-1} \left[ {\cal A}^L_{\mu} + \partial_\mu \right] U + \frac12  U \left[ {\cal A}^R_{\mu}  + \partial_\mu \right] U^{-1} \, , \cr
\hat {\cal A}_\mu (x) &=& \frac12 U^{-1} \left[ {\cal A}^L_{\mu} + \partial_\mu \right] U - \frac12  U \left[ {\cal A}^R_{\mu}  + \partial_\mu \right] U^{-1} \, , \cr
U(x) &=& e^{\frac{ i \pi(x)}{f_\pi}} \quad , \quad {\cal A}^{L(R)}_{\mu}(x) = {\cal V}_\mu (x) \pm {\cal A}_\mu (x) \, ,
\eeqa 
and the $\psi_n(z)$ modes satisfy  
\beqa
\kappa \int d z \, K_z^{-1/3} \psi_n (z) \psi_m ( z) =  \delta_{nm} \quad  ,  \quad 
- K_z^{1/3} \partial_{ z} \left[ K_z \partial_{z} \psi_n ( z) \right] = \lambda_n \,  
\psi_n (z) \, .
\eeqa

Using the Kaluza-Klein expansion (\ref{KKexpansion}) and integrating the $z$ coordinate we get a four-dimensional 
effective lagrangian of mesons and external $U(1)$ fields. 
The  vector (axial vector) mesons are represented by the fields $v_\mu^n (x)$ ($a_\mu^n (x)$) and 
correspond to the modes $\psi_{2n-1} (z)$ ($\psi_{2n} (z)$). The pion is represented by 
the field $\pi(x)$ and corresponds to the mode $\psi_0(z)$. In addition, we have external $U(1)$ vector 
(axial) fields represented by ${\cal V}_\mu$ (${\cal A}_\mu$). 

\medskip

In order to have a diagonal kinetic term, the vector mesons are redefined as 
$ \tilde v_\mu^n = v_\mu^n + (g_{v^n}/M_{v^n}^2) {\cal V}_\mu$  and the 
quadratic terms in the vector sector take the form \cite{Sakai:2005yt} :
\beqa
{\cal L}_2 &=& \frac12 \sum_n \left [ {\rm Tr} \left(\partial_\mu \tilde v_\nu^n - \partial_\nu \tilde v_\mu^n \right)^2 
+ 2  M_{v^n}^2 {\rm Tr}  \left(\tilde v_\mu^n - \frac{g_{v^n}}{M_{v^n}^2} {\cal V}_\mu \right)^2 \right ] \, , \nonumber
\eeqa
where
\beqa
 M_{v^n}^2 = \lambda_{\ss 2n-1} M^2_{\Lambda} \quad , \quad
 g_{v^n} = \kappa  \, M_{v^n}^2 \int d z \,K_z^{-1/3} \psi_{\ss 2n-1}(z) \, . \nonumber
\eeqa 

The mixed term $g_{v^n} \tilde v_\mu^n {\cal V}^\mu$ represents the  decay of the photon into vector mesons which is a 
holographic realization of  vector meson dominance.

\section{Form factors and structure functions of vector mesons in the Sakai-Sugimoto model}

In section 2 we described how the hadronic structure functions and form factors arise in the electromagnetic scattering of hadrons. 
The hadronic structure functions describe the inclusive interaction between a virtual photon and a hadron (Deep Inelastic Scattering).
They carry information of the partonic structure of hadron through the parton distribution functions. The latter are probability densities 
that describe how the partons (valence quarks, gluons and sea quark-antiquark pairs) distribute inside a hadron along the longitudinal momentum.
The electromagnetic form factors of hadrons describe the exclusive interaction between a photon and two hadrons. In the elastic case these quantities 
give information about the distribution of charge and magnetic moment inside a hadron. 
In the non-elastic case they describe the transition between different hadronic states and are usually written in terms of
helicity amplitudes. 
Below we describe how to calculate electromagnetic form factors and structure functions of vector mesons in the Sakai-Sugimoto model. 

\subsection{Electromagnetic form factors}

Electromagnetic form factors of vector mesons have been obtained previously using holographic bottom-up (phenomenological) models \cite{Grigoryan:2007vg,Grigoryan:2007my}
 as well as top-down string models dual to supersymmetric field theories \cite{Hong:2003jm,RodriguezGomez:2008zp}. Below we review the calculation of vector meson electromagnetic
form factors in the Sakai-Sugimoto model \cite{BallonBayona:2009ar} \footnote{ In ref. \cite{Bayona:2010bg} the vector meson electromagnetic form factors have been 
also calculated in a recently proposed string model for chiral symmetry breaking \cite{Kuperstein:2008cq}}. 

\medskip

From the cubic terms in the 4-d effective meson lagrangian  discussed in the previous section we can extract the vector meson interaction term
\beqa
 {\cal L}_{vvv} &=& \sum_{n, \ell, m}   g_{v^n v^\ell v^m} {\rm Tr} \Big\{  \left(\partial^\mu {\tilde v}^{\nu\, n} - \partial^\nu {\tilde v}^{\mu \, n} \right) 
 [{\tilde v}_\mu^\ell, {\tilde v}_\nu^m] \Big \}  
\eeqa
where $g_{v^n v^\ell v^m}$ are 4-d effective couplings given by the integral 
\beqa
g_{v^n v^\ell v^m} = \kappa \,  \int d  z \,K_z^{-1/3} 
\psi_{2n-1}(z) \psi_{2\ell -1}(z) \psi_{2m-1}( z) \,.
\eeqa

\begin{figure}[!ht]
\begin{center}
\includegraphics[height=3.5cm,width=4cm]{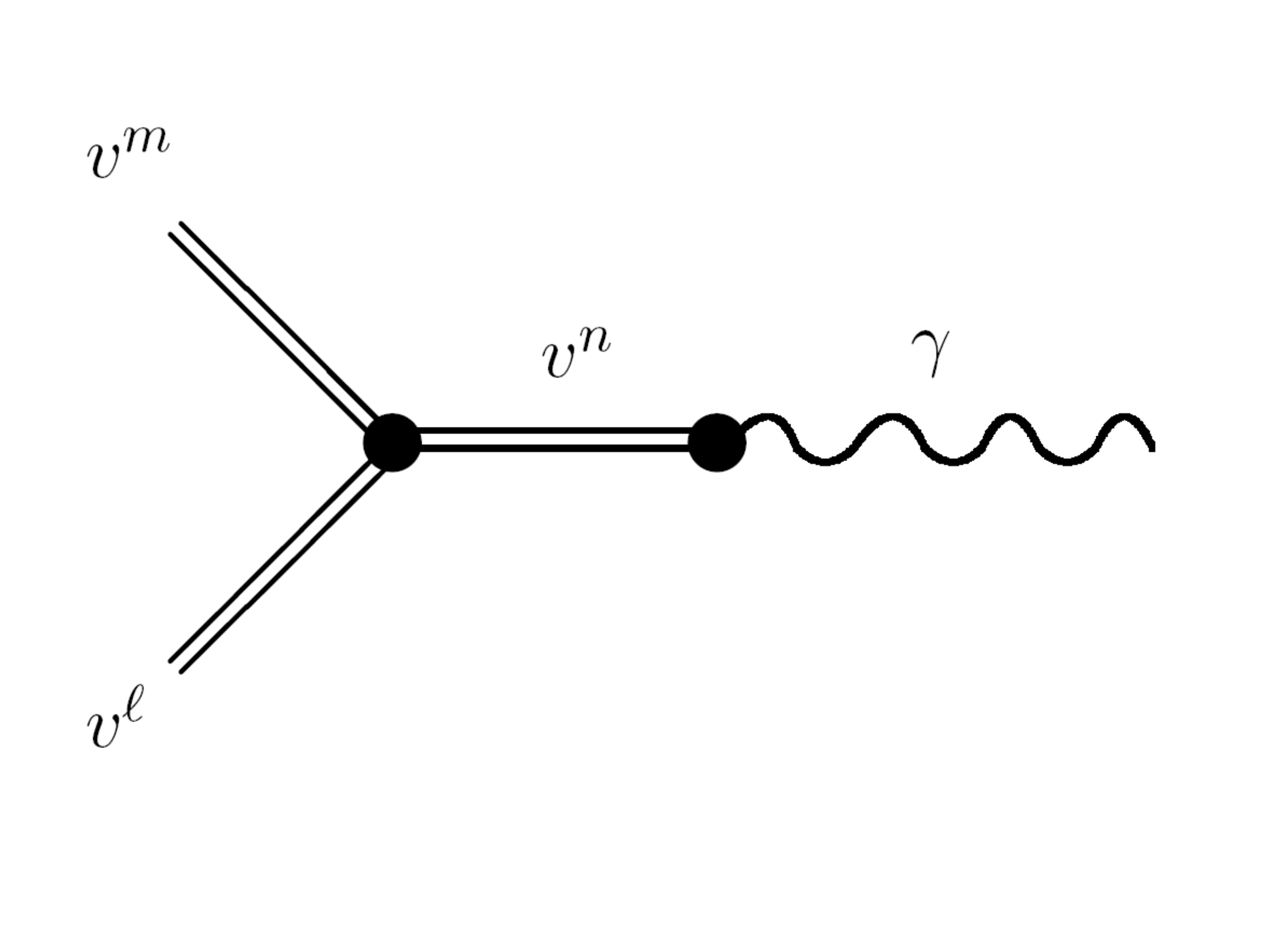}
\caption{Exclusive scattering of a photon and two vector mesons}
\label{FeynFormFactor}
\end{center}
\end{figure}

Now consider the exclusive scaterring of a photon with two vector mesons $v^m$ and $v^\ell$ as described in Figure \ref{FeynFormFactor}.
The photon decays into a vector meson $v^n$ that propagates and interact with the external vector mesons. 
Using the Feynman rules associated to this process  we find the electromagnetic current matrix element 
for vector mesons states :
\begin{eqnarray}
&&\langle v^{m}(p), \epsilon \vert J^{\mu }(0) \vert v^{\ell }(p+q), \epsilon' \rangle 
= \left[\sum_{n=1}^\infty {g_{v^n}g_{v^mv^nv^\ell}} \Delta^{\mu \sigma} (q, m_n^2) \right ] \cr 
&&\epsilon^\nu {\epsilon'}^\rho \Big[ \eta_{\sigma\nu}(q-p)_\rho  + \eta_{\nu\rho}(2p+q)_\sigma 
-  \eta_{\rho\sigma}(p + 2q)_\nu \Big] 
\end{eqnarray}
where $\Delta^{\mu \sigma} (q, m_n^2)$ is the propagator of a massive vector particle with momentum $q$ and mass $m_n^2$. Using the holographic sum rule
\begin{equation}
\sum_{n=1}^\infty \frac{g_{v^n}g_{v^nv^mv^\ell}}{M_{v^n}^2} = \delta_{m\ell}
\label{sumrule}
\end{equation}
and the transversality of the initial and final polarizations we get the simple expression 
\begin{eqnarray}
\langle v^{m}(p), \epsilon \vert J^{\mu }(0) \vert v^{\ell }(p+q), \epsilon' \rangle  &=& \epsilon^\nu {\epsilon'}^\rho \Big [ \eta_{\nu\rho}(2p+q)_\sigma  
+ 2(\eta_{\sigma\nu} q_\rho - \eta_{\rho\sigma}q_\nu) \Big] \cr 
&\times&  \left( {\eta^{\mu\sigma} - \frac{q^\mu q^\sigma}{q^2}} \right) {\cal F}_{v^m v^\ell}(q^2) \,,
\label{Formlong2}
 \end{eqnarray}
where 
\begin{equation}
{\cal F}_{v^m v^\ell}(q^2)=\sum_{n=1}^\infty \frac{g_{v^n}g_{v^nv^mv^\ell}}{q^2+M_{v^n}^2} 
\label{form:vn}
\end{equation} 
is the generalized vector meson form factor. The elastic $\rho$ meson form factors can be extracted from the case  $m = \ell =1$  :
\begin{eqnarray}
\langle v^{1}(p), \epsilon \vert J^{\mu }(0) \vert v^{1 }(p + q), \epsilon' \rangle &=& \Big \{ (\epsilon \cdot \epsilon') (2p+q)^\mu  \cr
&+& 2\left[ \epsilon^\mu (\epsilon'\cdot q) - {\epsilon'}^\mu (\epsilon \cdot q) \right] \Big \}
{\cal F}_{v^1 v^1}(q^2) \,, 
\label{Formelastic}
 \end{eqnarray} 
Comparing this result with the expansion (\ref{FormDecomp}) we find the $\rho$ meson form factors:
\begin{equation}
F_1(q^2) = F_2(q^2) =  {\cal F}_{v^1 v^1}(q^2) \,\,,\,\, F_3(q^2) = 0 \,.
\end{equation}

\begin{figure}[!ht]
\begin{center}
\includegraphics[height=3cm,width=4.5cm]{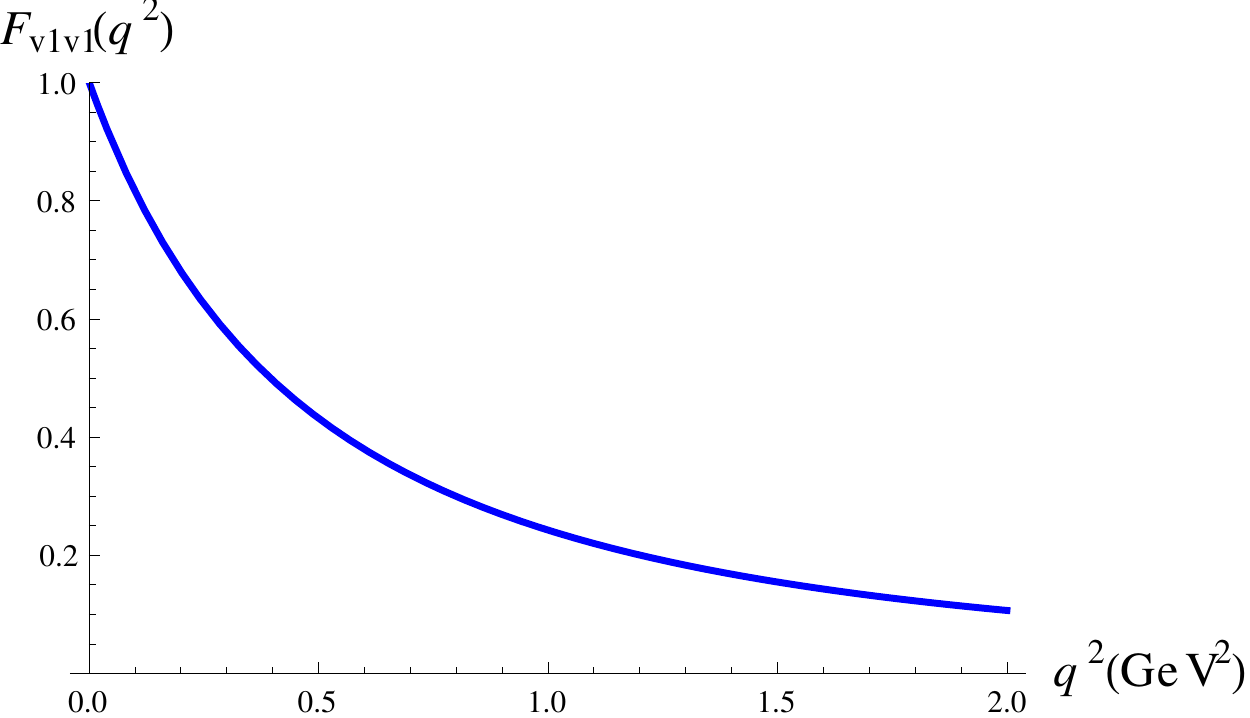}
\caption{The $\rho$ meson form factor 
in the Sakai-Sugimoto model.}
\label{formfactorrho}
\end{center}
\end{figure}

In Figure \ref{formfactorrho} we plot the $\rho$ meson form factor ${\cal F}_{v^1 v^1}(q^2)$. The form factor begins at ${\cal F}_{v^1 v^1}(0) =1$ (as expected from charge normalization)  and  
decreases approximately as $q^{-4}$ for large $q^2$ in accordance with other non-perturbative approaches to QCD (see for instance \cite{Ioffe:1982qb}) .

\medskip 

In order to extract some static properties of the $\rho$ meson it is useful to define the  electric, magnetic, and quadrupole form factors as
\beqa
&& G_E = F_1 - \frac{q^2}{6 M^2} \Big[ F_2 - ( 1 + \frac{q^2}{4 M^2})  F_3 \Big]
\quad , \quad 
 G_M = F_1 + F_2\,,
\cr  
&& G_Q = - F_2 + \left( 1 + \frac{q^2}{4 M^2} \right) F_3 \,.
\eeqa 

In the Sakai-Sugimoto model and for the $\rho$ meson they take the form
\begin{equation}
G_E = ( 1  + \frac{q^2}{6 M^2} ) F_1 \quad , \quad 
G_M = 2 F_1 \quad , \quad 
G_Q = - F_1  \,.
\end{equation}

From these form factors, we can extract the $\rho$ meson electric radius 
\begin{equation}
\langle r^2_{\rho}\rangle = -6\frac{\mathrm{d}}{\mathrm{d}q^2}G_E(q^2)|_{q^2=0} = 0.5739 \, ,
\end{equation}
and the magnetic and quadrupole moments
\begin{equation}
\mu = \frac{1}{2 M} G_M(q^2)|_{q^2=0} = \frac{1}{M} \quad , \quad 
D \, \equiv \,  \frac{1}{M^2} G_Q(q^2)|_{q^2=0} = - \frac{1}{M^2} \;.
\end{equation}

\subsection{Structure functions}   

Deep Inelastic Scattering in AdS/QCD was first investigated in a bottom-up model for the case of scalar particles \cite{Polchinski:2002jw}. 
Further development in bottom-up and top-down models include the large $x$ regime  \cite{BallonBayona:2007qr,BallonBayona:2008zi,Pire:2008zf,Bayona:2011xj,Koile:2011aa} 
as well as the small $x$ regime where Pomeron exchange dominates \cite{Brower:2006ea,Hatta:2007he,BallonBayona:2007rs,Cornalba:2008sp,Cornalba:2009ax,Brower:2010wf}. Here we review the calculation of structure functions 
in the Sakai-Sugimoto model  for the case where the initial state is a $\rho$ meson and the final state is one vector meson resonance \cite{BallonBayona:2010ae}. 

\begin{figure}[!ht]
\begin{center}
\includegraphics[height=4cm,width=5cm]{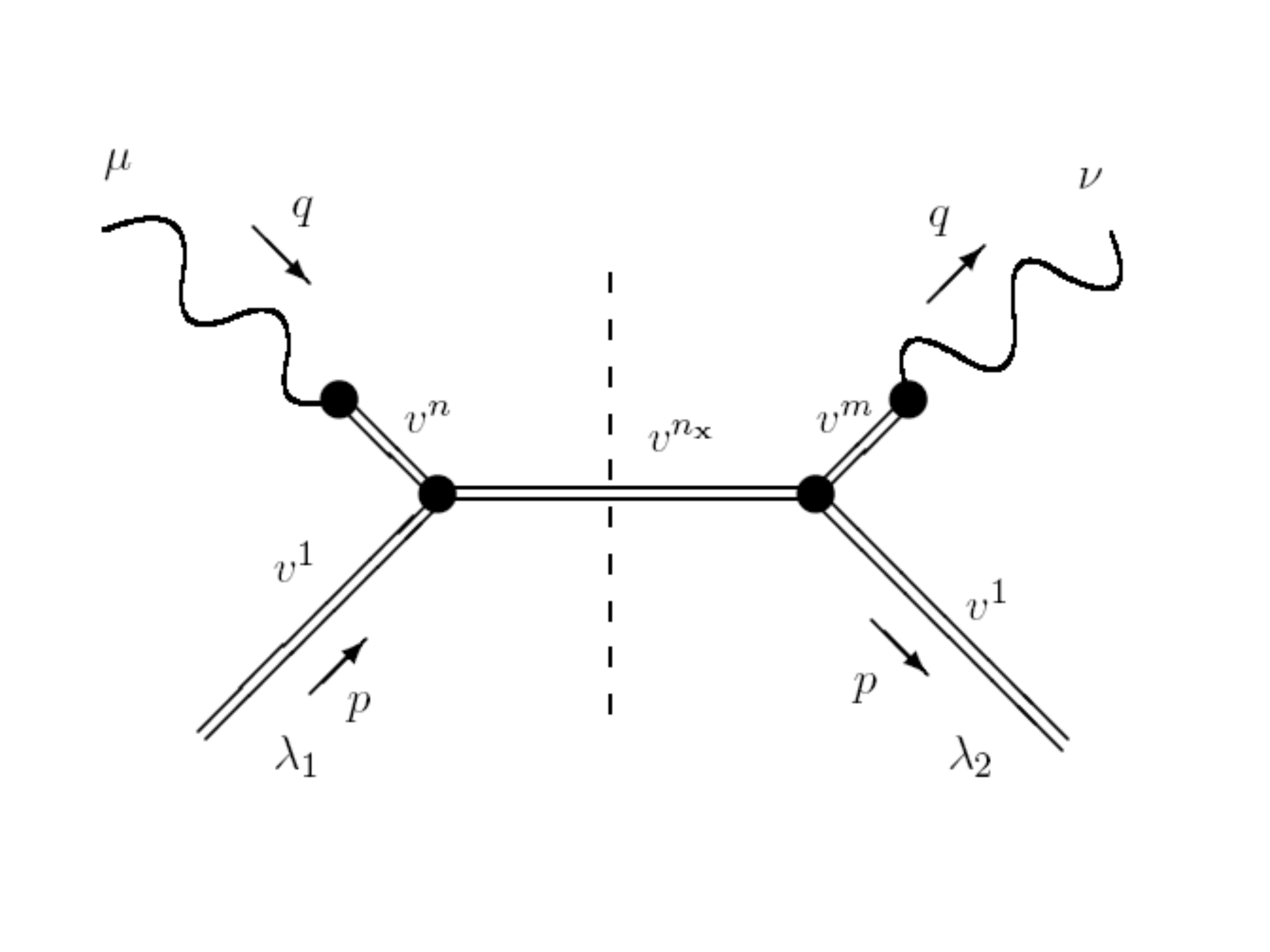}
\caption{Diagram for a tree level contribution to Compton forward scattering}
\label{FeynCompForw}
\end{center}
\end{figure}

The optical theorem relates the hadronic tensor to the imaginary part of the  tensor associated to the Compton forward scattering. 
Using the Feynman rules corresponding to the diagram  of Figure \ref{FeynCompForw} and the polarization vector identity 
\begin{equation}
\sum_\epsilon \epsilon^{\lambda_1} \epsilon^{\lambda_2} = \eta^{\lambda_1 \lambda_2} +
\frac{ p^{\lambda_1} p^{\lambda_2} }{M_{v^1}^2} \,,
\end{equation}
we obtain
\begin{eqnarray}
{\rm Im} T^{\mu\nu}&=&   \frac{N_f}{3}  
\left[\eta^{\mu \sigma_1} - \frac{q^{\mu} q^{\sigma_1}}{q^2}\right] 
\left[\eta^{\nu \sigma_2} - \frac{q^{\nu} q^{\sigma_2}}{q^2}\right] \cr \cr
&\times& \left[\eta^{\lambda_1 \lambda_2} + \frac{ p^{\lambda_1} p^{\lambda_2} }{M_{v^1}^2}\right]
\left[\eta^{ \rho_1 \rho_2} + \frac{(p+q)^{\rho_1} (p+q)^{\rho_2}}{W^2}\right] \cr 
&\times& \left[ \eta_{\sigma_1 \lambda_1} (2q)_{\rho_1} + \eta_{ \lambda_1 \rho_1} (2p)_{\sigma_1} +
\eta_{\sigma_1 \rho_1} (-2q)_{\lambda_1} \right] \cr 
&\times& \left[ \eta_{\sigma_2 \lambda_2} (2q)_{\rho_2} + \eta_{ \lambda_2 \rho_2} (2p)_{\sigma_2} +
\eta_{\sigma_2 \rho_2} (-2q)_{\lambda_2} \right] \cr
 &\times &
\sum_{n_{\bf x}} \left[{\cal F}_{v^1 v^{n_{\bf x}}} (q^2)\right]^2
 (2\pi) \delta [ M_{v^{n_x}}^2 - W^2 ] \, ,
\label{ImTmunu2}
\end{eqnarray}

\noindent where ${\cal F}_{v^1 v^{n_{\bf x}}} (q^2)$ is given in eq.(\ref{form:vn}). We can approximate the sum over the delta functions by an integral 
\begin{eqnarray}
\sum_{n_{\bf x}} \delta [ M_{v^{n_x}}^2 - W^2 ] 
&\equiv& \sum_{n_x} \delta [ M_{v^{n_x}}^2 -M_{v^{\bar n}}^2 ] 
= \int dn_x 
\left[\left| \frac{\partial M_{v^{n_x}}^2}{\partial n_x} \right|\right]^{-1}
\delta(n_x-\bar n)\cr
&\equiv& f(\bar n)\,.
\end{eqnarray}

The structure functions, defined in eq.(\ref{HadTensor}), take the form
\begin{eqnarray}
F_1(x,q^2)  &=&  \frac{4N_f}{3}  f(\bar n)  
 \Big[ {\cal F}_{ v^1 v^{\bar n}}  (q^2) \Big]^2 q^2 \left[ 2 + \frac{q^2}{4x^2 M_{v^1}^2} + \frac{ q^2} {W^2 x^2} ( x - \frac 12)^2 \right]
\cr
F_2(x,q^2)  &=&  \frac{4 N_f}{3}  f(\bar n) 
 \Big[ {\cal F}_{ v^1 v^{\bar n}}  (q^2) \Big]^2 \frac{q^2}{2x}  \left[ 3 + \frac{q^2}{ M_{v^1}^2} + \frac{ (q^2)^2} {M_{v^1}^2 W^2 x^2} 
( x - \frac 12)^2 \right] \,. \cr
&& \label{strutfunct}
\end{eqnarray}

\begin{figure}[!ht]
\begin{center}
\includegraphics[height=3cm,width=4cm]{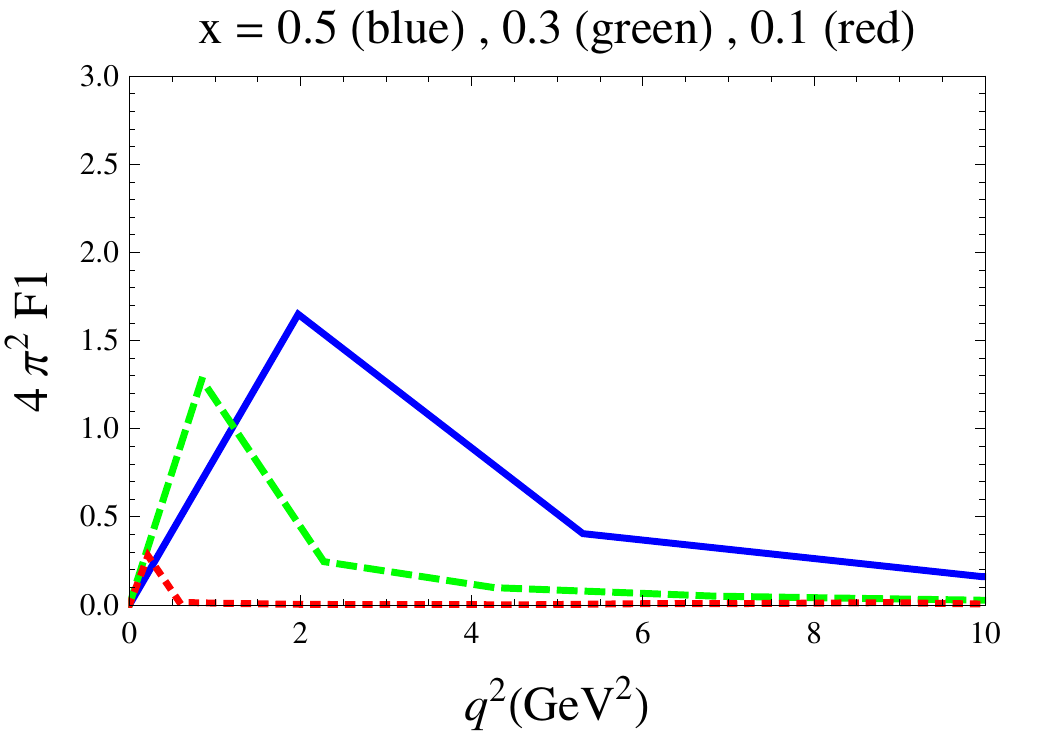}
\includegraphics[height=3cm,width=4cm]{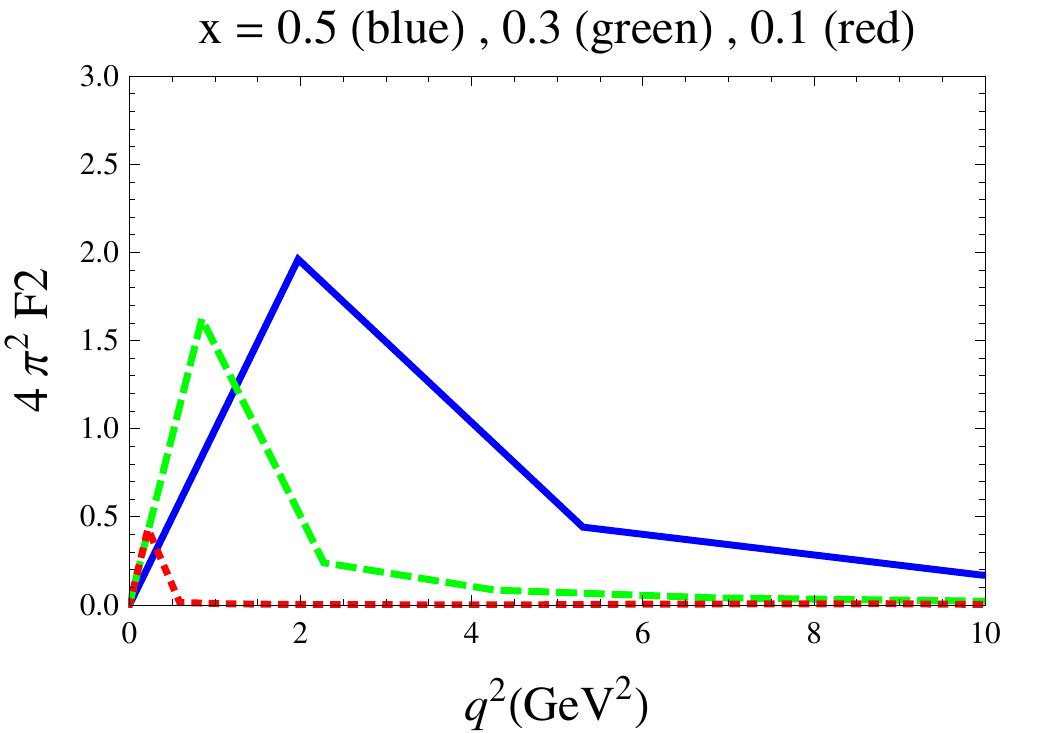}
\vspace{-0.3cm}
\caption{Structure functions as a function of the virtuality $q^2$}
\label{Stfuncq2}
\end{center}
\end{figure}

\begin{figure}[!ht]
\begin{center}
\includegraphics[height=3cm,width=4cm]{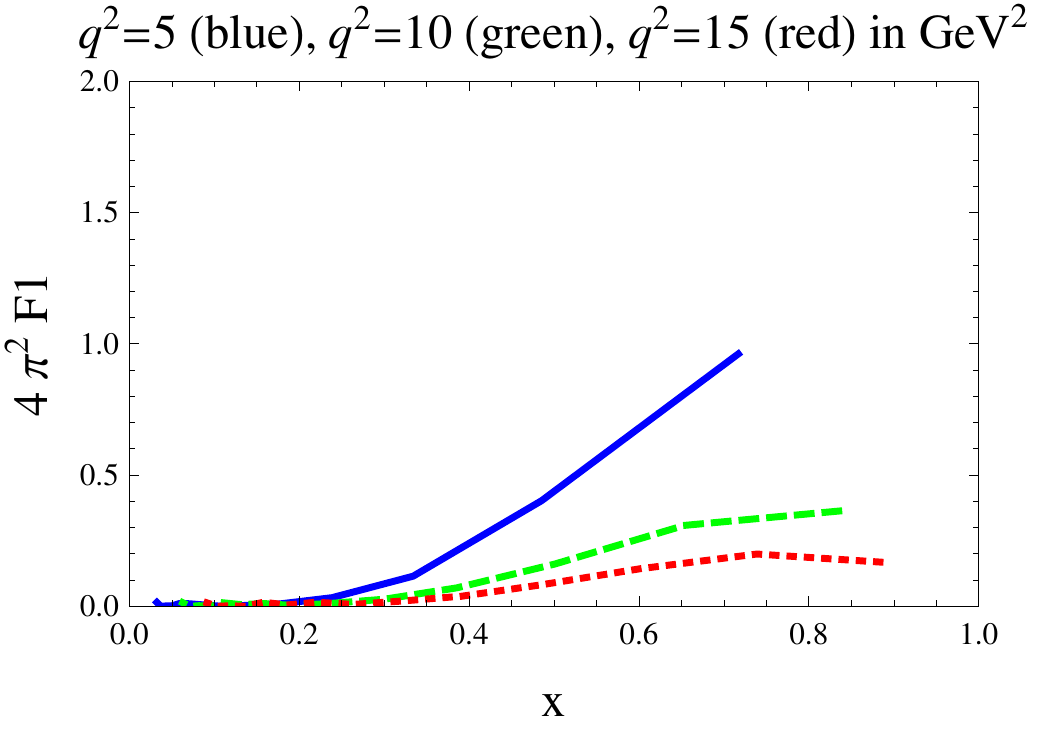}
\includegraphics[height=3cm,width=4cm]{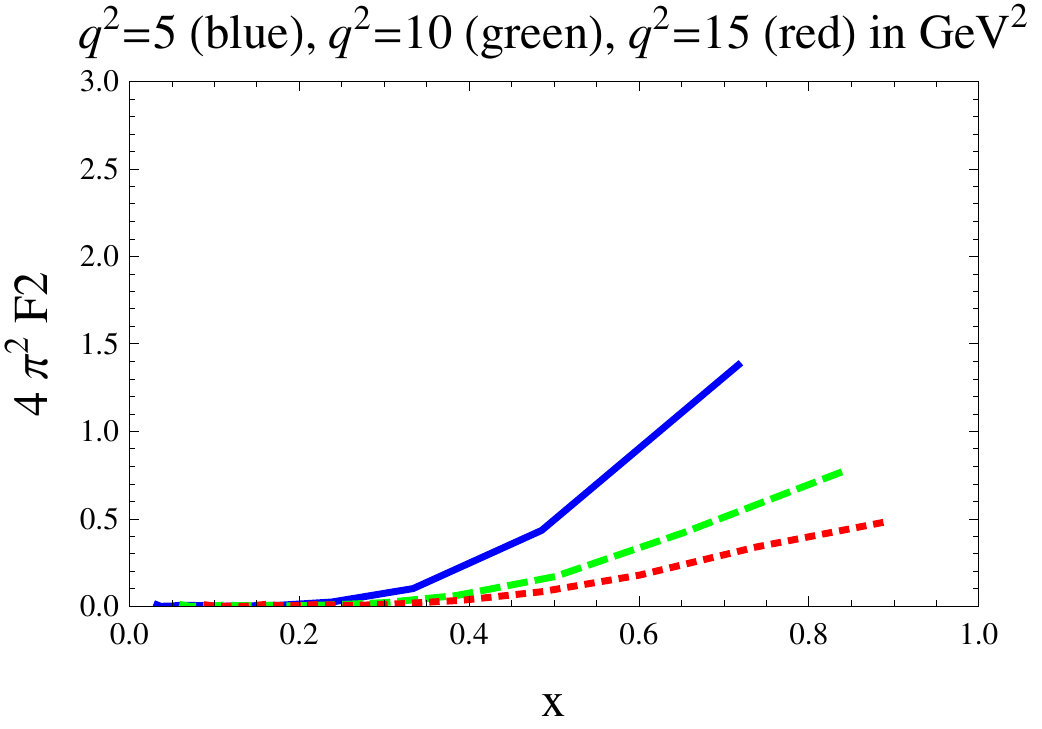}
\includegraphics[height=3cm,width=4cm]{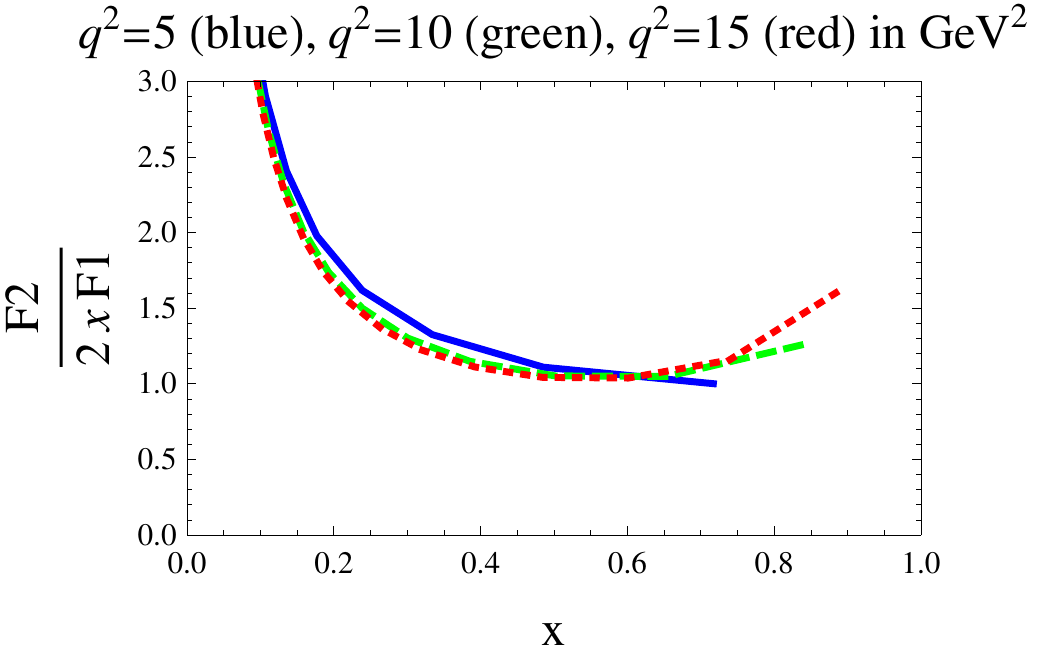}
\vspace{-0.3cm}
\caption{Structure functions as a function of the Bjorken variable $x$}
\label{Stfuncx}
\end{center}
\end{figure}

In figures \ref{Stfuncq2} and \ref{Stfuncx} we show the numerical results of the $\rho$ meson structure functions in the Sakai-Sugimoto model 
for $N_f = 1$. The plots show  $F_1$ and $F_2$  as a function of the virtuality $q^2$ 
and the Bjorken variable $x$. As expected the structure functions go to zero as $q^2$ goes to zero. Interestingly, the $x$ dependence of the structure 
functions suggest a valence quark behaviour and near $x=0.5$ we find an approximate Callan-Gross relation $F_2 \approx 2 x F_1$. However, in the approximation that we have considered here where the final state is 
composed by (only) one vector meson resonance the structure functions are very small. Further corrections may involve working with higher 
spin particles and $1/N_c$ corrections.  These issues will be investigated in future works.

\vspace{-1cm}

\section{Conclusions}

The Sakai-Sugimoto model offers a new approach for the  non-perturbative regime of hadronic interactions. In particular, it realizes in a simple way the property of vector meson dominance. 
In this proceeding we reviewed the calculation of electromagnetic form factors and structure functions of vector mesons in this model.
We considered only processes where one particle is produced in the final state. Processes involving two or more particle states arise as corrections of order $1/N_c$ and 
will be the subject of future research.

\medskip 

The perturbative QCD approach to hadron scattering involves factorization between  hard parton scattering and soft parton distribution functions. 
It would be interesting to understand this factorization in the AdS/QCD approach. Recently interesting models in AdS/QCD have been proposed
 to study the regime of high energies  ($x \ll 1 $) but low momentum transfer ($q^2 \ll \Lambda$) where the soft pomeron is relevant. To develop these
models in the context of the Sakai-Sugimoto model is another direction of future research.

\medskip

{\bf Acknowledgment: } Talk presented by C.A.B.B. at the Eleventh Workshop on Non-Perturbative Quantum Chromodynamics, Paris, June 6-10, 2011 (eConf C1106064). 
He thanks the organizers for the hospitality and the stimulating atmosphere. This work was sponsored by the STFC Rolling Grant ST/G000433/1 (United Kingdom)
and the grants from CNPq, Capes and Faperj (Brazil).

\end{document}